# Using Graph Theory and a Plenoptic Sensor to Recognize Phase Distortions of a Laser Beam

CHENSHENG WU,[1,*] JONATHAN KO,[1] CHRISTOPHER C. DAVIS[1]

[1]*University of Maryland, College Park, MD 20742, USA*
*\*Corresponding author: wuchensheng07@gmail.com*

**Atmospheric turbulence causes fluctuations in the local refractive index of air that accumulatively disturb a wave's phase and amplitude distribution as it propagates. This impairs the effective range of laser weapons as well as the performance of free space optical (FSO) communication systems. Adaptive optics (AO) can be applied to effectively correct wavefront distortions in weak turbulence situations. However, in strong or deep turbulence, where scintillation and beam breakup are common phenomena, traditional wavefront sensing techniques such as the use of Shack-Hartmann sensors lead to incorrect results. Consequently, the performance of AO systems will be greatly compromised. We propose a new approach that can determine the major phase distortions in a beam instantaneously and guide an AO device to compensate for the phase distortion in a few iterations. In our approach, we use a plenoptic wavefront sensor to image the distorted beam into its 4D phase space. A fast reconstruction algorithm based on graph theory is applied to recognize the phase distortion of a laser beam and command the AO device to perform phase compensation. As a result, we show in our experiments that an arbitrary phase distortion with peak to peak value up to $22\pi$ can be corrected within a few iteration steps. Scintillation and branch point problems are smartly avoided by the plenoptic sensor and its fast reconstruction algorithm. In this article, we will demonstrate the function of the plenoptic sensor, the fast reconstruction algorithm as well as the beam correction improvements when our approach is applied to an AO system.**

*OCIS codes: (010.1330) Atmospheric turbulence; (010.1080) Active or adaptive optics; (010.7350) Wave-front sensing; (070.7345) Wave propagation; (100.5070) Phase retrieval.*

http:

## 1. Introduction

Adaptive optics is the art of instantly sensing and imposing patterns on the complex amplitude of a beam wave to counteract the effects of turbulence on beam propagation. It is often used to deblur the images from telescopes in astronomy where the incident wavefront gets disturbed by the earth's atmosphere[1] at the end of its propagation path. An adaptive optics system often requires certain forms of feedback such as a live view of the beam or its wavefront[2]. Sometimes a retro-reflector[3] or a beacon laser[4] is mounted at the receiver site as a reference for correction, so that informative feedback information can be obtained at the transmitter site and convenient algorithms can be applied to perform beam corrections. In general, the type of feedback determines the correction algorithm used in AO systems. For example, image metrics are image based evaluations for the iterative movements of an AO system[5].

It is not surprising to find that improving sensing technology has a fundamental influence on adaptive optics systems[6]. For example, by adding holographic filters in front of photodetectors, one can recover the basic wavefront information[7] at decent speed, which shows high potential to improve the performance of adaptive optics devices in beam correction. In fact, advances in wavefront sensing are continually strengthening the correction powers of traditional AO systems.

In this paper, we introduce a new approach by using a modified plenoptic camera as a plenoptic wavefront sensor to handle beam distortion in strong or deep turbulence. In principle, the modified plenoptic sensor originates from the light field camera[8,9] which keeps the light field in a high resolution image and is flexible for refocusing at various depths. By making modifications to the optical components of a traditional light field camera, one can use it to solve coherent light field problems[10]. With the retrieved coherent light field, our corresponding algorithm can extract information to instruct each actuator in an AO device (such as a deformable mirror) to correct the phase distortion of the beam. Consequently, the number of iteration steps can be significantly reduced when compared with conventional beam correction methods such as the SPGD algorithm[11]. In fact, our approach uses the plenoptic sensor to guide the AO system to achieve major phase correction of a distorted beam within a few iteration steps, and further optimizes the result by the SPGD algorithm. Thus, with little compromise on the overall operation speed of a conventional AO system, we can speed up the correction speed by a minimum factor of 10.

Scintillation is also a disrupting factor in adaptive optics[12]. It is recognized as quick oscillations of local intensity values on the sensor that jeopardize the wavefront sensing process, making the sensing device partially or completely "blind" to the situation. However, with the plenoptic sensor, scintillation effects are suppressed for two distinct reasons: light field separation in the hardware layer and selective reconstruction algorithm in the software layer. In other words, scintillation problems can be handled by the plenoptic sensor and its reconstruction algorithm to a large extent.

In this article, Part 2 shows the plenoptic sensor and its reconstruction algorithm. Part 3 demonstrates the experimental results to verify our approach. Conclusions and discussion are contained in Part 4.

## 2. Plenoptic Sensor and Fast Reconstruction Algorithm

### A. Plenoptic sensor

The plenoptic sensor is designed based on the structure of a light field camera. A light field camera trades the resolution of a 2D image for a more general 4D representation of ray vectors. Each illuminated pixel can be translated into a uniform congruence of rays with corresponding geometrical and angular information, while the square root of intensity represents the field strength. Once the light field is retrieved, it can be back traced and form an image that is differently focused. In other words, the image of a light field camera contains adequate information to reconstruct a group of pictures that have different focal depth. However the light field camera can't be used to observe distorted laser beams, because coherence of the beam will generate interference patterns that can't be regarded as true representations of field strength.

We modified the optical arrangement in a light field camera to accomplish the task of recognizing the complex amplitude of a coherent light field[13]. A compact structure diagram of the plenoptic sensor is shown in Fig.1:

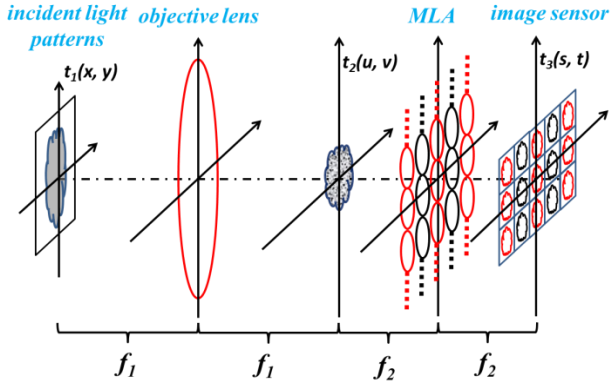

Fig. 1. Diagram of plenoptic sensor

In the diagram, we use a micro-lens array (MLA) with focal length $f_2$ to form a mini "Keplerian" telescope array with shared objective lens with focal length $f_1$. In addition, we regulate the diameter of the objective lens $(D)$ and the width of a MLA cell $(d_0)$ so that they satisfy the condition:

$$\frac{D}{f_1} \leq \frac{d_0}{f_2}. \quad (1)$$

In order to maximize the observation of the light field, we match the f/# of the objective lens and the MLA cell. Geometric and wave analysis of the plenoptic sensor[14] can be used to serve as a light field analyzer that sorts the complex pattern of the incident beam onto an array of image cells that are much simpler to solve[14]. In fact, the cell indexes (M, N) represent the first order of wavefront tip and tilt, while the geometrical distribution in each image cell is a linear map of geometries in the real world.

The 2-step wave propagation equations with regard to the shared focal plane and the plane of image sensor are expressed respectively as[14]:

$$t_2(u,v) = \frac{1}{j\lambda f_1} \int\!\!\!\int_{-\infty}^{+\infty} t_1(x,y)\exp\left(-j\frac{2\pi}{\lambda f_1}(xu+yv)\right)dxdy. \quad (2)$$

$$t_3(s,t) = \frac{1}{j\lambda f_2}\iint t_2(u,v)\sum_{M,N} rect\left(\frac{u+s-2Md_0}{d_0}\right)\cdot$$
$$rect\left(\frac{v+t-2Nd_0}{d_0}\right)e^{-j\frac{2\pi}{\lambda f_2}[(u-Md_0)(s-Md_0)+(v-Nd_0)(t-Nd_0)]}dudv. \quad (3)$$

In equation (2), $\lambda$ is the wavelength of the beam, $t_1(x, y)$ is the complex amplitude of the field at the front focal plane of the objective lens and $t_2(u, v)$ is the complex amplitude of the field when $t_1(x, y)$ propagates to the back focal plane of the objective lens. We assume that there is no turbulence distortion inside the plenoptic sensor and therefore $t_1(x, y)$ can be regarded as the instantaneous distorted beam. Equation (2) is also a fundamental conclusion from Fourier optics that in paraxial approximation, fields at the front and back focal planes of a convex lens are Fourier transforms of each other[15]. More specifically, by defining:

$$f_x = \frac{u}{\lambda f_1}, \qquad f_y = \frac{v}{\lambda f_1}. \quad (4)$$

Thus $t_2(u, v)$ is a scaled Fourier transform of field $t_1(x, y)$. Equation (3) describes the process in which the field $t_2(u, v)$ is subsampled by the MLA cells and processed by a second layer of local Fourier transforms into field $t_3(s, t)$ to form an image on the final focal plane array. In equation (3), the function rect(*) represents the square aperture of the MLA cell with width $d$.

Intuitively, a distorted beam is separated into an array of image cells that have narrow ranges of angular acceptance and are mutually exclusive in the 4D space of light rays (2D in geometrics plus 2D in momentum) Therefore, in strong or deep turbulence, where the beam breaks up into a number of interfering patches (small wavelets that act like planewaves), the plenoptic sensor can sort their angular spectrum in a linear order and map them onto corresponding image cells. In other words, given the same distorted beam, it is much harder to find interfering patches compared with methods that sense the beam without angular spectrum separation. In fact, two patches have to be close to each in both geometric and angular parameters to generate an interference pattern.

Since the geometric and angular information of a patch can be retrieved by the coordinates of corresponding illuminated pixels, algorithms can be developed to back trace those patches into the field to virtually reconstruct a phase screen[16] to represent the momentary cause of distortion in the channel. AO corrections can be applied based on the reconstruction results. Since the turbulent channel changes rapidly, the reconstruction must also be instantaneous and relatively accurate. In practice, a fast and practical reconstruction algorithm is preferred over delicate but time consuming reconstruction approaches.

## B. Connections between wavefront sensing and AO systems

We first show how the reconstruction result can be applied to guide AO systems in correcting a distorted beam through turbulence. Assume that the AO device used for beam correction has finite controllable elements (generally called "channels" or "actuators"), while the actual turbulence has more degrees of freedom. Thus each channel has to be used independently in order to maximize the set of applicable deformations for beam correction. Intuitively, the process can be expressed with the following equations:

$$\Phi_{turbulence}(x_i, y_i; T) = f_i\{I(s,t), s, t; T\}. \quad (5)$$

$$D_{actuator}(x_j, y_j; T+\Delta t) = g_j\{\Phi_{turbulence}(x_i, y_i; T)\}. \quad (6)$$

Where $f_i\{*\}$ is the function for the $i^{th}$ independent spot on the reconstructed phase screen and $g_j\{*\}$ is a linear function for the $j^{th}$ independent channel on an AO device used for correction. $I(s,t)$ is the intensity values of the plenoptic image, with $(s,t)$ representing the pixel coordinates. It is clear that in order to achieve the maximum degree of correction, the cardinality of the set of functions $\{f_i\}$ must be no less than the cardinality of the set $\{g_j\}$. Otherwise, we can find some cases of phase distortions within the correcting capacity of the system that will fail due to insufficient reconstruction. $T$ is the timestamp of the current plenoptic image, and $\Delta t$ is the time difference between acquiring the current frame and forming a correction order.

Without loss of generality, we restrict the distortion to be phase distortion. Amplitude distortions are generally less severe than phase distortions and can be directly detected and compensated[17]. If the phase screen is reconstructed on the surface of a deformable mirror (named as the "correction plane" in the following discussion) and each recovered spot geometrically corresponds to one of the actuators in the device, equation (6) can be explicitly expressed as:

$$g_j = -\frac{1}{2}\delta_{ij} f_i. \quad (7)$$

We have used Einstein notation in equation (7). This means exact "compensation" since the reconstructed phase screen and the correction plane are point to point matched.

In more general situations, the reconstructed phase screen and the correction plane are two separated planes in the channel. But each should be able to form perfect "compensation" when virtually propagated to the plane of the other. The process can be illustrated by Fig.2:

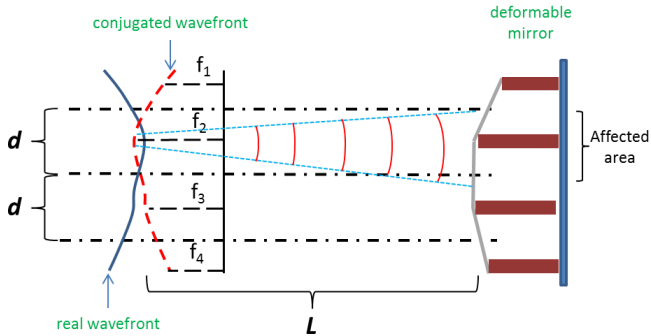

Fig. 2. Correction diagram based on reconstruction

In Fig.2, $d$ is the spacing between neighboring actuators, and L is the distance between the reconstructed phase screen and the correction plane. We assume that the layout of $g_j$ is identical with the layout of $f_i$. A rule of thumb is to form the correction command with the following equation:

$$g_i \approx -\frac{1}{2}\delta_{ij} f_j + \frac{1}{2}\overline{f_i - f_j}\{f_j, f_i \mid d_{ij} \leq \frac{0.61\lambda L}{d}, j \neq i\}. \quad (8)$$

In equation (8), $d_{ij}$ is the distance between actuators with index $i$ and $j$. It is inferred from equation (8) that an averaging aperture that is not diffraction limited can be used to account for the minor changes in local gradient. Intuitively, the maximum distance between the reconstructed phase screen and the correction plane should satisfy:

$$L_{max} \approx 0.4\frac{D^2}{\lambda}. \quad (9)$$

In equation (9), $D$ is the diameter of the correction plane. Note that $L_{max}$ can be regarded as an upper range limit for this pre-distortion strategy to work. For example, if the system has an aperture of 0.1m, the upper range of predistortion strategy is estimated as 4 km. And beyond this range, it is meaningless to perform any wavefront sensing and apply predistortion to the transmitted beam. A tighter bound can be derived by considering the actual level of atmospheric turbulence.

## C. Fast reconstruction algorithm

Without loss of generality, we set the goal of phase reconstruction to satisfy the minimum requirement: we only need to recover the phase information of $N$ points, which has the identical geometric distribution of the $N$ channels of the AO device in use. Then based on the previous discussion, we can virtually form an exact compensation on the reconstructed phase screen and propagate to the correction plane to determine the deformation needed on the AO device. An extension of our fast reconstruction algorithm is possible if more resolution points are needed, which is covered at the end of this section.

The fast reconstruction is established based on graph theory, so that the actual AO device in use and the light field information provided by our plenoptic sensor can be jointly considered. In fact, a digraph[18] is used for each reconstruction, with its vertices representing individual actuators of the AO device and its directional edges carrying phase difference between vertices. A reconstruction is completed if all the phases of the vertices are known. Intuitively, if there are N vertices in use, only N-1 edges are required to form a spanning tree to connect all the vertices together and reveal their phases.

Because the neighboring actuators in an AO device are usually equally spaced, we only make use of the shortest edges that join nearest pairs of vertices together. For example, in a Cartesian layout of actuators where neighboring elements are equally spaced, all the edges used to form a spanning tree will have the same length. Thus the total number of edges to be retrieved from a plenoptic image can be expressed as:

$$|E| = \frac{1}{2}\sum_{i=1}^{N} d_i. \quad (10)$$

In equation (10), $d_i$ is the number of nearest vertices of the $i^{th}$ vertex, and $|E|$ denote the cardinality of the set of edges. The spanning tree with N-1 edges is a subset of the edges in equation (10), denoted as set $\{E\}$.

We show next how to select the most informative edges to obtain the spanning tree, and therefore retrieve the phases on the vertices. For simplicity, a "dummy" AO system with only seven channels of control is used in our explanation. Its actuators form a hexagonal layout as shown in Fig.3:

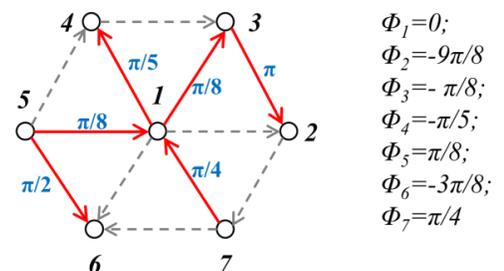

Fig. 3. Simple example of the reconstruction digraph

In Fig.3, the directional red edges represent the selected edges of the spanning tree. The numbered vertices are the geometric locations of the actuators of the device. The gray dashed edges are not selected by the fast reconstruction algorithm. The phase information retrieved from the plenoptic sensor is marked in pair with each corresponding edge. Thus, with the selected spanning tree, the phases on the vertices are deterministic.

A normal plenoptic image and the edge selection algorithm for the spanning tree is illustrated in Fig.4:

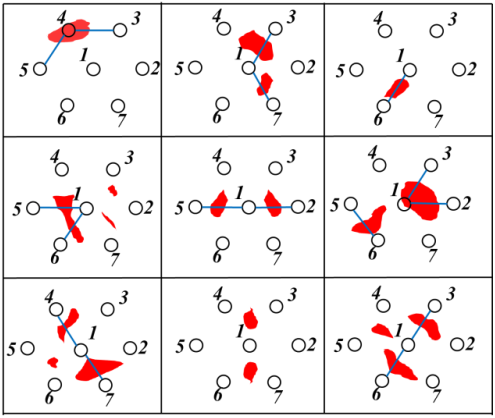

Fig. 4. Edge selection process on the plenoptic image for maximum spanning tree on the digraph presented by Fig. 3.

In Fig.4, the solid black lines mark the division of image cells and the red islands are the images of the light patches when the distorted wavefront is imaged by the plenoptic sensor. We virtually make copies of the layout of the vertices of the "dummy" AO into each image cell to show their corresponding locations. Based on the previous discussion, there are 12 edges in the graph: $E_{\{1,2\}}$, $E_{\{1,3\}}$, $E_{\{1,4\}}$, $E_{\{1,5\}}$, $E_{\{1,6\}}$, $E_{\{1,7\}}$, $E_{\{2,3\}}$, $E_{\{3,4\}}$, $E_{\{4,5\}}$, $E_{\{5,6\}}$, $E_{\{6,7\}}$. The direction of these edges can be arbitrarily assigned. We then use a box counting method to sum up (over all $M$ and $N$ indexes) the pixel values covering each edge and rank them in descending order. Using the summed-up intensity as the weight for each edge, the edges selection process is converted into a maximum spanning tree problem[19]. The "Greedy" method[20] can be used to practically determine the structure of the tree with the following steps:
(1) Start with a graph with all the vertices and zero edge.
(2) Take the first element in the edge set E (the edge with highest weight) and put it into the graph.
(3) If the edge doesn't form a circuit in the graph, keep it.
(4) Delete the selected edge from the edge set E.
(5) Go back to step (2) until N-1 edges are selected.

Once the structure of the maximum spanning tree is determined, the phase information of the edges can be calculated by the following equation:

$$\varphi_{E_{\{j,k\}}} = \frac{\sum_{(s,t)\in E_{\{j,k\}}} I(s,t)\vec{v}_{E\{j,k\}} \cdot \left(\frac{Md_0}{f_1}, \frac{Nd_0}{f_1}\right) \cdot \frac{2\pi l_0}{\lambda}}{\sum_{(s,t)\in E_{\{j,k\}}} I(s,t)}. \quad (11)$$

In equation (11), $I(s, t)$ is the pixel value with coordinates $(s, t)$ on the plenoptic image. $(M, N)$ are the indexes for an image cell with width $d_0$. $\vec{v}_E$ is the unit directional vector of the edge and $l_0$ is the length of the edge. $\lambda$ is the wavelength and $f_1$ is the focal length of the plenoptic sensor's objective lens. $E_{\{j,k\}}$ is the edge between vertices $j$ and $k$. Then, by arbitrarily appointing any vertex to be the reference point $(\varphi=0)$, one can find the phases of other vertices by traversing the tree structure and using the phase difference information carried by its branches.

If the reconstruction requires more resolution points, one can use a denser set of vertices and go through the same procedure. Accordingly, the actual form of equation (6) should be determined to relate the higher resolution of reconstruction to the lower resolution of wavefront correction on the AO system.

### D. Suppressed scintillation effect and branch point problem

The scintillation effect in atmospheric turbulence is generally regarded as the fluctuations in intensity distribution when a beam is observed by the receiver[21]. In strong or deep turbulence conditions, a laser beam will typically break up into a group of small time varying patches. Scintillations result from the rapid change of patch geometries as well as their coherent interference[22] patterns. This will cause inaccuracies in wavefront reconstruction. For example, if two interfering patches enter the same MLA cell of a Shack-Hartmann sensor, the local wavefront can't be determined since there is more than one spot under the cell domain. The loss of phase continuity in wavefront reconstruction is commonly referred to as a branch point problem[23,24]. Similarly, dark regions caused by scintillation also lead to branch point problems on the Shack-Hartmann sensor.

It is evident that on the plenoptic sensor, scintillation due to coherent interference is largely reduced. Since two patches must overlap both in their geometry and momentum information to interfere with each other, interference is less likely to happen. In fact, suppose that the angular information of a patch is randomly distributed over n-by-n image cells, the probability of patch interference is roughly reduced by a factor of $n^2$ when compared to direct imaging of the beam.

Beam wander effects, on the other hand, are not suppressed in the plenoptic sensor images. It is not difficult to find that the location of a patch may jump from one image cell to another due to its angular variations. Even with perfectly suppressed interference, beam wander still causes certain level of scintillation on the plenoptic sensor. Our fast reconstruction algorithm, however, handles this situation in the software layer by dynamically selecting the most informative branches to form a reconstruction. In fact, since the patterns of the plenoptic image changes over time, the algorithm adaptively avoids dark pixels that provide little information about the distorted beam. Moreover, the maximum spanning tree will naturally avoid controversies in reconstruction results through different integral paths. In other words, it only generates one of all the possible paths to retrieve the phase of each vertex. Thus, branch point problems are cleverly avoided by our fast reconstruction algorithm.

Branch points can still be found when inspecting the discarded branches with a high intensity sum. Intuitively, those branches have high intensities and therefore carry informative information of phase difference between vertices. They are discarded simply because they don't contribute to the construction of the maximum spanning tree. When adding one of those discarded branches to the constructed spanning tree, a unique circuit will be formed. If the loop sum of phase changes around the circuit is not close to zero, we can determine that one or more branch points exist within the enclosed area of the circuit. In addition, since the algorithm only cares about pixels representing the geometry of the |E| edges, branch points resulting from other pixels on the image sensors are naturally ignored. Thus it justifies the statement that the branch point problem can't be removed, but is avoided by the algorithm.

In conclusion, the plenoptic sensor combined with its "adaptive" fast reconstruction algorithm is less affected by scintillation and branch point problems in performing wavefront reconstruction and can be applied to provide better correction strategies in AO systems.

## 3. Experimental results

### A. Device and experiment settings

We use an OKO deformable mirror (piezoelectric, 37 actuators, 30mm diameter, hexagonal layout) as the AO device to experimentally verify our algorithm. The layout of the deformable mirror (DM) is shown as in Fig.5:

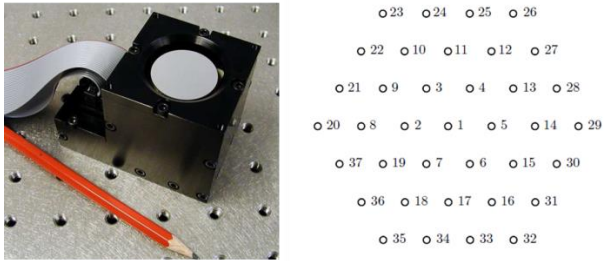

Fig. 5. Image and actuator layout of OKO PDM 37

Fig.5 shows that the 37 actuators in the DM are equally spaced in a hexagonal format. By applying equation *(10)*, there are 90 edges in the graph and 36 of them need to be selected by the algorithm to retrieve phases on the 37 vertices to form a correction command.

Two conditions have been defined in order to make better evaluation of our algorithm:
(1) We use the DM to generate an initial phase distortion, and then use the same DM to correct the beam.
(2) We integrated the SPGD algorithm in the platform so that it can be invoked at any time to take over the correction process.

It is obvious to see that condition (1) matches with the exact "compensation" condition as described by equation (7). Also we are assured that the device is capable to correct the generated phase distortion as the DM will be a "flat mirror" when the beam is corrected. Condition (2) provides the convenience to show the improvements to the algorithm by comparing it with an effective and conventional AO algorithm in correcting the same phase distortion.

The experiment diagram and picture of the instruments are shown as in Fig.6:

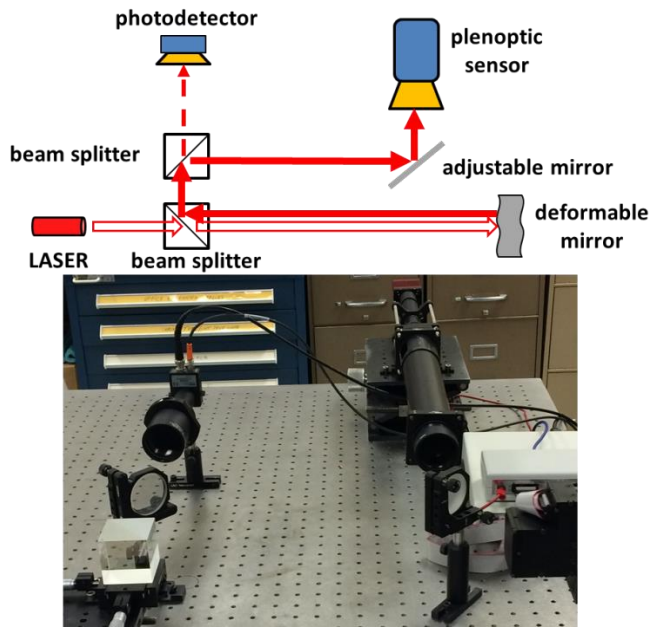

Fig. 6. Experimental diagram and set up

A 5mW 632.8nm He-Ne laser source (Thorlabs: HNL050L) is used and expanded to a collimated beam with 40mm diameter. The beam propagates through a 50mm cubic beam splitter and reflects off the surface of the DM. The reflected beam is truncated by the DM to an aperture with 30mm diameter and a superposed phase change. Then the distorted beam with 30mm diameter is directed to the plenoptic sensor to measure the phase distortions as shown by the solid red arrows in Fig.6. A secondary channel directs the distorted beam to a photodetector to measure the undistorted amount of optical intensities as shown by the red dashed arrow in Fig.6. This secondary channel uses an aspherical lens of focal length of 150mm and a pinhole with 0.1mm diameter placed at the back focal spot of the lens. A photodetector (Thorlabs: PDA100A) is put behind the pinhole. Thus, any significant distortion will deviate part of the beam away from the focal spot and reduce the intensity value on the photodetector.

The plenoptic sensor uses an Allied Vision GX1050 monochrome camera (1024×1024 in resolution with 5.5 μm pixel width, and operating at 109 fps) as its image sensor. The objective lens has a clear aperture of 50mm and focal length of 750mm. The MLA used is a 10mm×10mm microlens array with 300μm cell width and 5.1 mm focal length (Edmund Optics: #64-476). Limited by the size of the image sensor, the plenoptic sensor has 18×18 (324) as the maximum number of image cells.

When no distortion is applied to the beam, the photodetector obtains the highest intensity value and the plenoptic sensor shows a round beam in one image cell (defined as the center block). When a distortion is applied by the DM, the photodetector shows a drop of value and the plenoptic sensor shows distributed patches in different image cells. Therefore, the goal of SPGD correction (based on the feedback from the photodetector) is to maximize the intensity value on the photodetector. The goal of guided correction (based on the feedback from the plenoptic sensor) is to rehabilitate the image of the beam to the center block.

The image metric used on the plenoptic sensor is "power out of the bucket (POB)," which measures the intensity sum outside the center block divided by a constant scaling factor. The photodetector measures the power of the undistorted optic flux and is conventionally referred to as "power in the bucket (PIB)"[25]. We choose POB as our image metric for the plenoptic sensor because of its simplicity and its complementary property with the PIB measurement on the photodetector. Based on POB, one can implement an advanced image metric that varies significantly when some patches move further away from the center block.

To initiate the phase distortion, we use the DM to mimic various modes of Zernike Polynomials[26] with peak to peak value of 5.5λ (λ=632.8nm). This maximum deformation is experimentally determined, and a larger magnitude of deformation can't be accurately enforced by the DM due to hysteresis effects in the piezoelectric actuators. Note that the when the beam is reflected from the surface of the DM, the actual peak to peak phase distortion is <11λ. Since the Zernike Polynomials are Eigen modes of an arbitrary deformation on the DM, the expected iteration steps are bounded by the best and worst cases of correcting those Zernike modes.

To correct the phase distortion, we use an algorithm called "Guided SPGD" method, which first uses the plenoptic sensor to iteratively use our fast reconstruction algorithm to correct for the major phase distortion, and then start a conventional SPGD algorithm for optimization when the beam is almost corrected (lower than a threshold image metric value). For comparison, the "Pure SPGD" method is applied to correct the same deformation, which strictly uses SPGD algorithm for each iteration step of correction.

## B. Experiment results

We start with a simple tip/tilt case in Zernike Polynomials as shown by Fig.7:

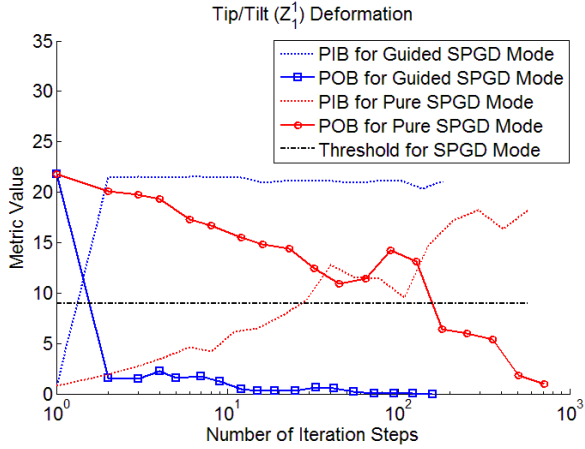

Fig. 7. Experimental result for correcting Tip/Tilt deformation

In Fig.7 a tip (or equivalently tilt) deformation is applied by the DM. The deformation can be expressed as:

$$Z_1^1(\rho,\theta) = A \cdot \rho \qquad 0 \leq \rho \leq 1, \ \theta \in [0, 2\pi). \qquad (12)$$

In equation (12), "A" is the magnitude of the phase deformation ($A=11\lambda$), while $\rho$ and $\theta$ represent the normalized radius and angle respectively.

As shown by the legend, the solid blue line represents the improving curve of the "Guided SPGD" method. It is obvious that the major correction is completed in one step through the guidance of the plenoptic sensor. Then the correction enters the SPGD mode to make tiny random iterations to minimize the image metric (or equivalently, maximize the power collected by the photodetector). At the 10th iteration step, the beam has already reached the vicinity of the optimum point. This result of "Guided SPGD" mode is also reflected by the "PIB" curve on the photodetector data, which is a complement for the "POB" curve. As shown by the solid red line, when the "Pure SPGD" method is applied to correct for the same distortion, it takes about 150 iteration steps to reach the threshold for the "Guided SPGD" method and many more iteration steps (about 800) to get close to the optimum point. The difference can be intuitively explained: when the tip/tilt deformation is applied, it moves the beam from one image cell to another on the plenoptic sensor. The movement is recorded by all the 90 edges and the spanning tree will have every branch indicating the phase tip/tilt (the deformation can be fully recognized regardless of which branches are selected). Thus the correction orders for the actuator on the DM will correct the deformation directly in the 1st step.

The "SPGD" method, on the other hand, will generate an identical and independent distribution (i.i.d) of Bernoulli $(p=0.5)$ over set $\{+\varepsilon, -\varepsilon\}$ for each actuator, where $\varepsilon$ is a small deformation. Then, it determines the gradient of the image metric over this small step and tells the next movement whether to either forward this step (if it decreases the "POB" image metric) or reverse this step (if it increases the "POB" image metric) in proportion to the magnitude of the gradient. Thus, improvement can be guaranteed in terms of every 2 iteration steps. In practice, we make slight improvements on the SPGD to make assured convergence in roughly every 1.5 steps by empirically equalizing the magnitude of the trial step $\varepsilon$ with the actual correcting deformation determined by SPGD. Note that we make $\varepsilon$ linearly scaled by the image metric value to accelerate the convergence (similar to the gain coefficient $\gamma$ in the formal theory of SPGD[11]). Then, the SPGD method will either keep the deformation or reverse the deformation to make assured decreasing in 1.5 steps (the trial step is actually the correction step). However, with 37 actuators, the convergence is relatively slow because the typical result of a random trial step makes little improvement on the image metric (statistically half of the actuator get closer to the correct deformation while the other half of the actuators get farther away from the correct deformation). Intuitively, for a DM with N actuators, each iteration step has $(1/2)^N$ chance to guess the perfect directions of movement for all the actuators. Therefore, given N=37, the chance that SPGD can generate significant improvement in the first step is almost zero. The plenoptic sensor, however, has the capability to recognize the major phase distortion and inform the actuators to move correctly not only in terms of directions, but also in terms of proper magnitude. Thus, the extra intelligence provided by the plenoptic sensor can dramatically reduce the overall steps needed for beam correction (or equivalently make convergence to optimization more quickly).

We should also account for the fact that the image based feedback loop takes longer per iteration than the time required for a photodetector feedback loop. For example, in our experiment, the image based feedback loop operates at 65 fps (9.2ms for image acquisition, 1.5ms for DM control and 4.6ms for computer process time including calculating the image metric). Comparatively, the photodetector feedback loop operates at 482 fps (1.5ms for DM control and 0.57ms for photodetector data acquisition plus computer process time). Therefore, by regarding each guidance step from the plenoptic sensor as about 8 iterations in SPGD mode, we can find the corresponding improvement in terms of time. For the tip/tilt deformation, since there is only 1 step guidance from the plenoptic sensor, the improvement by "Guided SPGD" method is still significant when compared with "Pure SPGD" method. We also stress that these execution times are currently constrained by working in a Microsoft Windows environment with its associated overhead, and the max operational speed of our current deformable mirror. Execution times per iteration can be dramatically improved by implementing FPGA operation with a faster response deformable mirror. More results are shown as in Fig.8 -13:

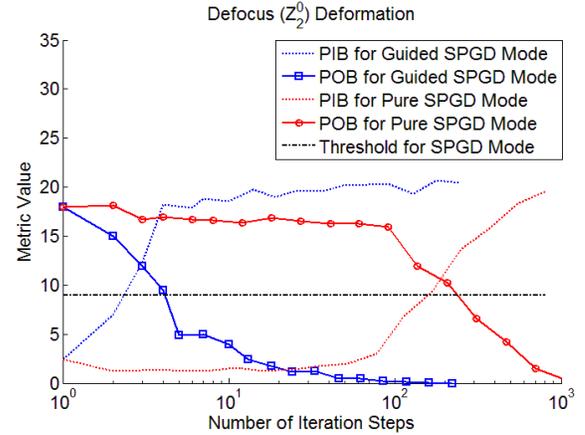

Fig. 8. Experimental result for correcting Defocus deformation

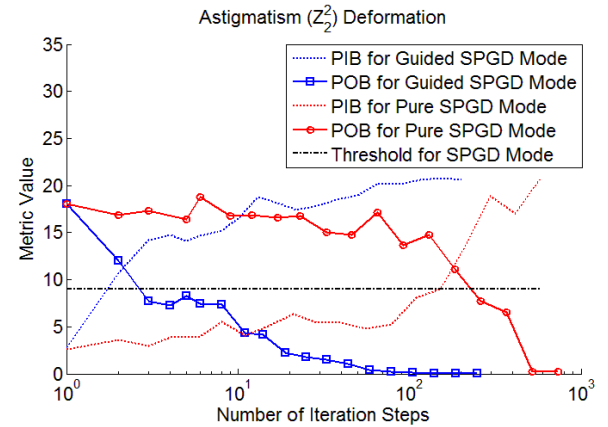

Fig. 9. Experimental result for correcting Astigmatism deformation

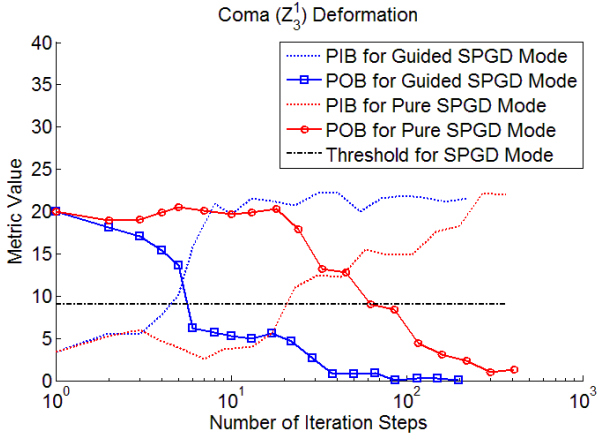

Fig. 10. Experimental result for correcting Coma deformation

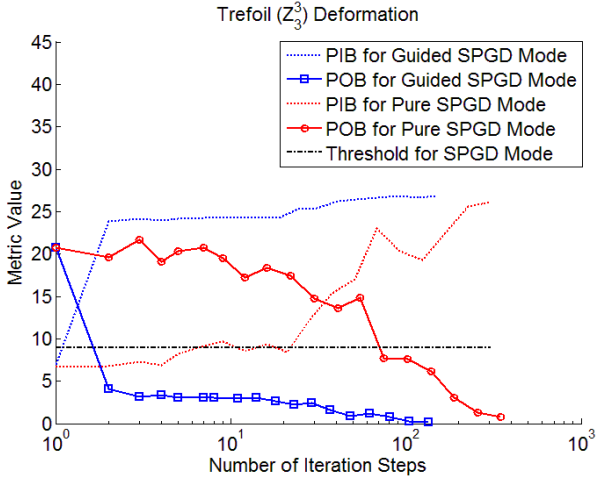

Fig. 11. Experimental result for correcting Trefoil deformation

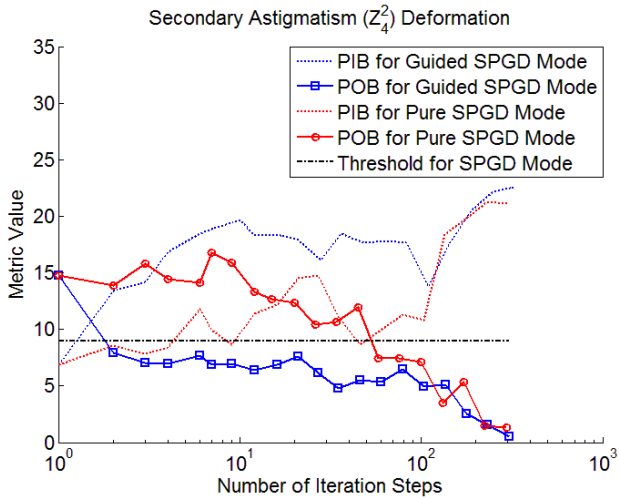

Fig. 12. Experimental result for correcting Secondary Astigmatism deformation

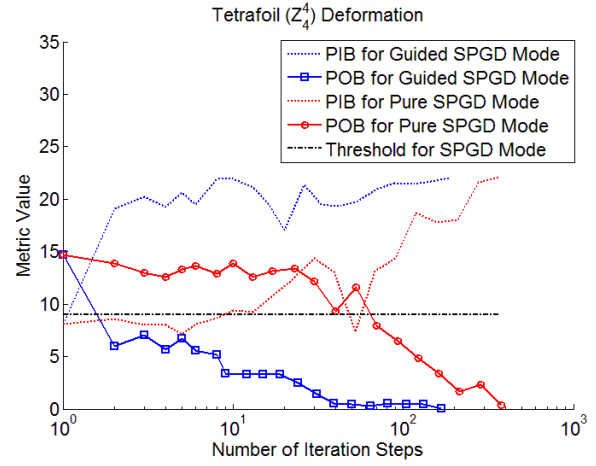

Fig. 13. Experimental result for correcting Tetrafoil deformation

Fig.7-13 show the comparison between the "Guided SPGD" method and "Pure SPGD" method for different orthogonal modes that are achievable on the DM. Given "A" as the magnitude of the phase deformation ($A=11\lambda$), $\rho$ and $\theta$ as the normalized radius and angle. We can express those deformations as:

$$\text{Defocus}: \quad Z_2^0(\rho,\theta) = A \cdot \rho^2. \quad (13)$$

$$\text{Astigmatism}: \quad Z_2^2(\rho,\theta) = \frac{A}{2}[\rho^2 \cdot \cos(2\theta)+1]. \quad (14)$$

$$\text{Coma}: \quad Z_3^3(\rho,\theta) = \frac{A}{2}[(3\rho^3-2\rho) \cdot \cos(4\theta)+1]. \quad (15)$$

$$\text{Trefoil}: \quad Z_3^3(\rho,\theta) = \frac{A}{2}[\rho^3 \cdot \cos(3\theta)+1]. \quad (16)$$

$\text{Secondary Astigmatism}:$

$$Z_4^2(\rho,\theta) = \frac{A}{2}[(4\rho^4-3\rho^2) \cdot \cos(4\theta)+1]. \quad (17)$$

$$\text{Tetrafoil}: \quad Z_4^4(\rho,\theta) = \frac{A}{2}[\rho^4 \cdot \cos(4\theta)+1]. \quad (18)$$

The parameters from equation (13) to equation (18) should satisfy:

$$0 \leq \rho \leq 1, \quad \theta \in [0, 2\pi). \quad (19)$$

As shown by Figs. 8-13, the "Guided SPGD" method converges to optimized phase correction much more effectively than the "Pure SPGD" method. The blue "PIB" curve (provided by the photodetector) for the "Guided SPGD" method in each deformation shows the exponentially increasing intensity on the receiver. This is also reflected by the blue "POB" curve for the "Guided SPGD" method in each deformation. In comparison, note that the red "POB" curve for "Pure SPGD" in each deformation is generally a concave curve with regard to the logarithm of iteration steps. This reveals that SPGD is more efficient in correcting weak distortion than in correcting strong distortion. Equivalently, the same conclusion can be drawn by inspecting the convexity of the red "PIB" for the "Pure SPGD" method in each deformation. Intuitively, the major proportion of iteration steps is conducted by SPGD in both methods, while the "Guided SPGD" uses guidance from the plenoptic sensor to find a much better starting point for SPGD than the unguided "Pure SPGD" method. It costs very few steps to find the starting point with the help of the plenoptic sensor, and therefore, much faster convergence can be achieved by the "Guided SPGD" approach.

A detailed result of the fast reconstruction algorithm can be explained in Fig.14 by using the Trefoil deformation as an example. The upper-left plot shows the plenoptic image when the Trefoil deformation is applied (the non-illuminated image cells are not shown in this plenoptic image). The upper-right plot is the maximum spanning tree that the algorithm derives to recognize the major phase distortion pattern. The bottom-left plot shows the original command on each actuator of the DM to generate the deformation. The bottom-right plot shows the reconstructed phase for the actuators by using the fast reconstruction algorithm. The maximum spanning tree indicates which set of pixels the fast reconstruction should use to form a "sub-optimal" reconstruction of the phases on the vertices (optimal reconstruction can be achieved by using all the pixels). In other words, the fast reconstruction algorithm makes use of the pixels that represent the location of the *N-1* edges rather than the whole 2D plenoptic image, which brings down the input data requirement from 2D to 1D. It is clear that the fast reconstruction recognizes the dominant phase distortion correctly. Then the one step phase compensation will be sufficient to fix most of the phase distortion in this "Trefoil" case. This explains the dramatic drop of image metric value in Fig.11 after the first step correction using the "Guided SPGD" method.

phase deformation to the collimated beam and makes the beam diverge over a range of angles, the plenoptic image sees all kinds of patches in different blocks. The fast reconstruction algorithm picks up the outside patches (since they are largest and most illuminated compared with other patches) and uses them to reconstruct the phases on the actuator to form a phase compensation suggestion for the next iteration. The upper-right image shows the result of first step correction. Clearly, the large and bright patches in the outside image cells are eliminated and packed into inner image cells. Then, similar corrections are formed iteratively in the second iteration (bottom left image), third iteration (bottom middle image) and fourth iteration (bottom right image). After the fourth guided correction step, the defocus is almost compensated as most of the optic flux is directed back to the center image cell. If the image metric drops below the threshold value of 9, the fast reconstruction will not see significant phase changes between vertices and we can invoke the SPGD method to make further optimization with a few iterations.

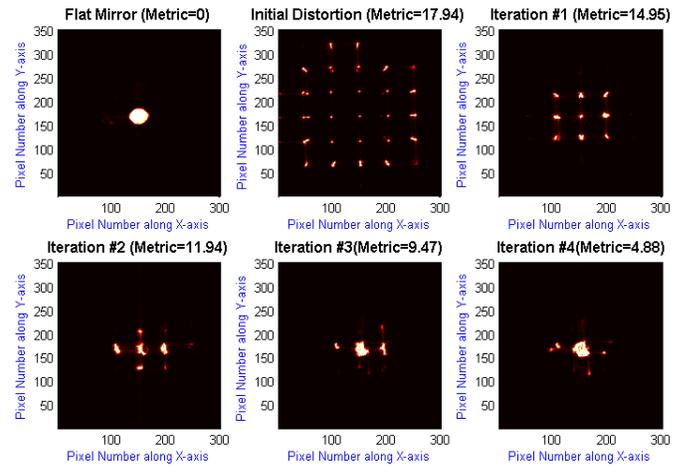

Fig. 15. Plenoptic images for each guided correction for "Defocus" deformation case (4 guided correction steps were taken)

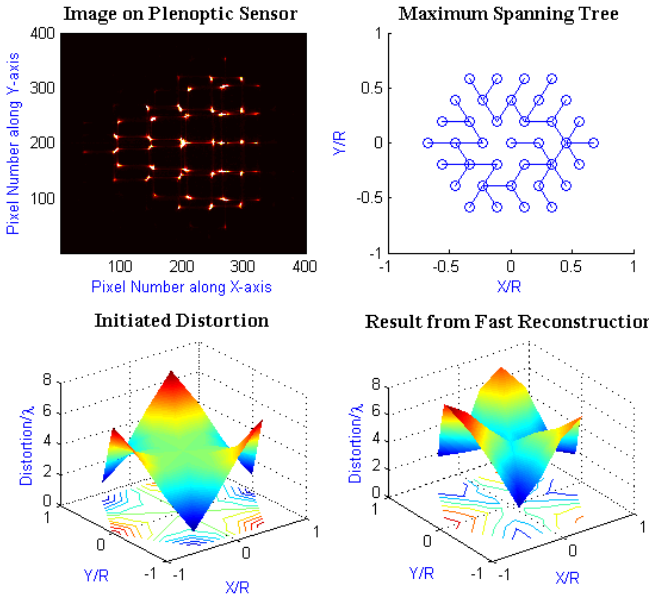

Fig. 14. Detailed result for a single step fast reconstruction by using the case "Trefoil" deformation

For some cases such as "Defocus" in Fig.8 and "Coma" in Fig.10, the "Guided SPGD" needs more than one step to accomplish the major correction (bring the image metric value down to the threshold value of "9" to trigger SPGD mode for optimization). This means that one step fast reconstruction does not suffice to recognize all the major phase distortions. However, the algorithm will iteratively recognize the most "significant" phase distortion patterns and suppress them with phase compensation. Intuitively, the word "significant" means the light patches that illuminated the pixels of certain edges most. The plenoptic image sequence for the first 4 guided correction steps in the "Defocus" case (Fig.8) are shown by Fig.15:

The step-by-step plenoptic images of the guided corrections show how the initial 4 correction steps by the plenoptic sensor can effectively make a distorted beam converge to an almost corrected form and trigger SPGD to perform the optimization. The upper-left image shows the collimated beam on the plenoptic sensor. The upper-middle image shows how the deformation distorts the beam on the plenoptic sensor. Since "Defocus" deformation introduces a hyperbolic

The image metric values over the first 10 iteration steps with the "Guided SPGD" method for all the applicable Zernike modes (as shown in Fig.7~13) on the DM are presented in Fig.16.

It can be concluded from Fig.16 that the major phase correction can be done by the plenoptic sensor and its fast reconstruction algorithm within few iteration steps (N≤5 in our experiments). Whenever the image metric value drops below the threshold value (M=9), the SPGD optimization process will be triggered to fix the remaining weak distortions in phase.

### C. Further Discussion and Comparison

Since SPGD utilizes random guesses (+ε or –ε, where ε is an arbitrary and small iteration step size) on each actuator of the AO device, the actual number of iteration steps fluctuates a lot to achieve correction for the same distortion. Thus it can be argued as "unfair" to compare the "Guided SPGD" and "Pure SPGD" methods by plotting outcomes from a single experiment since the "Pure SPGD" may get unlucky and result in a long convergence sequence. To rule out "bad luck" we repeated 100 trials using the "Pure SPGD" method to correct each Zernike Polynomial phase distortion case and selected the best correction case that requires minimum steps to converge. The minimum step is denoted by an integer $n_0$, where starting with $n>n_0$, the normalized variation of intensity on the photodetector is less than 2% (equivalently, the image metric value is smaller than 2). The results are shown in table 1. The correction uses the "Pure SPGD" method where the metric is the photodetector readouts.

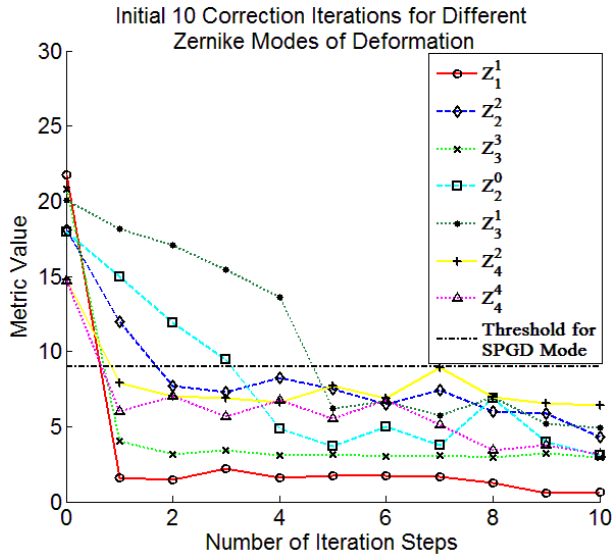

Fig. 16. First 10 corrections steps for each deformation case by using "Guided SPGD" method

**Table 1. Minimum iteration steps using "Pure SPGD" method (best result through 100 repeated corrections in each case)**

| Zernike Polynomial Modes | Minimum Number of SPGD Iterations |
|---|---|
| Tip/Tilt ($Z_1^1$) | 199 |
| Defocus ($Z_2^0$) | 238 |
| Astigmatism ($Z_2^2$) | 162 |
| Coma ($Z_3^1$) | 214 |
| Trefoil ($Z_3^3$) | 240 |
| Secondary Astigmatism ($Z_4^2$) | 304 |
| Diamond ($Z_4^4$) | 210 |

It is easy to find that in the case of a 37-channel AO device, the best "Pure SPGD" results cost around 200 iteration steps. Those results still can't beat the "Guided SPGD" method without best case selection. The presented "Guided SPGD" corrections typically completes at around 100 iterations (a case of exemption is shown by Fig. 12 with some "bad luck", since most of the corrections are still carried by "SPGD" algorithm). Thus, it is evident that the guided steps strongly speed up the convergence of correction.

In fact, the philosophy behind the improvement is that we have implemented a wavefront sensor that can extract the major phase distortion within the channel. Therefore, with deterministic direction and magnitude instructions for individual actuators in the AO device, the phase distortion is significantly suppressed within few iteration steps (for example, ≤5 in our experiments). For the remaining weak distortion, the plenoptic sensor can't provide further guidance since the center block will become the most illuminated cell, and SPGD can be invoked to complete the correction (perform optimization).

## 4. Conclusions

In this article, we have demonstrated the mechanism of our plenoptic sensor and its fast reconstruction algorithm using graph theory. The plenoptic sensor serves as a light field analyzer that separates the incident beam into patches whose geometry and wavefront tilt (with regard to the optic axis) is deterministic. In the graph, the vertices correspond to the individual control channels in the AO device and the edges are the extracted phase difference from the plenoptic image. A "greedy" method based on the most illuminated patches is invoked to form an intensity based maximum spanning tree to retrieve phases on the vertices. Then the AO device can be used to compensate the phases on the discrete vertices. Consequently, the compensation helps to remove the major phase distortion. By using the sensing-reconstructing-compensating process iteratively, an arbitrary phase distortion can be effectively corrected within few steps. The algorithm is fast because it uses a small portion of pixels to accomplish the recognition of major phase distortions.

In the experiments, we verified that the fast reconstruction algorithm can effectively recognize and instruct the AO device to perform phase compensation for large and arbitrary phase distortions. Consequently, our proposed "Guided SPGD" method achieves a much faster convergence speed than conventional "SPGD" methods. Judging by the speed to correct 90% percent of the phase distortion (image metric value drops below 9), the "Guided SPGD" method completes the work faster than the "Pure SPGD" method by a minimum factor of 10. Beam break up phenomenon in strong or deep turbulence can be effectively countered by the hardware structure of the plenoptic sensor and scintillation effects are adaptively avoided by the fast reconstruction algorithm that dynamically forms a maximum spanning tree based on the most informative patches.

In application, our plenoptic sensor and the fast reconstruction algorithm can be used to instruct conventional AO systems to form intelligent corrections (no random guesses). Also, in the field of adaptive beam combining, a huge challenge is that when the number of combined laser sources grows larger, convergence requires longer steps[27]. Our approach has the potential to provide solutions to the convergence problem in adaptive beam combining by treating the discrete laser sources as vertices in our graph theory based fast reconstruction algorithm.

## Funding Information

## References

1. Hardy, John W. Adaptive Optics for Astronomical Telescopes. Oxford University Press, 1998.
2. Liang, Junzhong, David R. Williams, and Donald T. Miller. "Supernormal vision and high-resolution retinal imaging through adaptive optics." JOSA A 14.11 (1997): 2884-2892.
3. Zhang, Yan, et al. "Adaptive optics parallel spectral domain optical coherence tomography for imaging the living retina." Optics Express 13.12 (2005): 4792-4811.
4. Tyson, Robert. Principles of Adaptive Optics. CRC Press, 2010.
5. Gladysz, Szymon, Natalia Yaitskova, and Julian C. Christou. "Statistics of intensity in adaptive-optics images and their usefulness for detection and photometry of exoplanets." JOSA A 27.11 (2010): A64-A75.
6. Roddier, Franccçois. "Curvature sensing and compensation: a new concept in adaptive optics." Applied Optics 27.7 (1988): 1223-1225.
7. Zepp, Andreas. "Holographic wavefront sensing with spatial light modulator in context of horizontal light propagation." In SPIE Remote Sensing, pp. 85350I-85350I. International Society for Optics and Photonics, 2012.
8. Ng, Ren, et al. "Light field photography with a hand-held plenoptic camera." Computer Science Technical Report CSTR 2.11 (2005).
9. Lumsdaine, Andrew, and Todor Georgiev. "The focused plenoptic camera." Computational Photography (ICCP), 2009 IEEE International Conference on. IEEE, 2009.
10. Wu, Chensheng, Jonathan Ko, William Nelson, and Christopher C. Davis, "Phase and Amplitude Wave Front Sensing and Reconstruction with a Modified Plenoptic Camera" SPIE Proceedings 9224, Laser Communication and Propagation through the Atmosphere and Oceans III, G-1 - G-13, 2014.
11. Liu, Ling, and Mikhail A. Vorontsov. "Phase-locking of tiled fiber array using SPGD feedback controller." In Optics & Photonics 2005, pp. 58950P-58950P. International Society for Optics and Photonics, 2005.
12. Barchers, Jeffrey D., David L. Fried, and Donald J. Link. "Evaluation of the performance of Hartmann sensors in strong scintillation." Applied optics 41, no. 6 (2002): 1012-1021.

13. Wu, Chensheng, and Christopher C. Davis. "Modified plenoptic camera for phase and amplitude wavefront sensing." SPIE Optical Engineering + Applications. International Society for Optics and Photonics, 2013.
14. Chensheng Wu, Jonathan Ko, and Christopher C. Davis, "Determining the phase and amplitude distortion of a wavefront using a plenoptic sensor," J. Opt. Soc. Am. A 32, 964-978 (2015).
15. Goodman, Joseph W. Introduction to Fourier optics. Vol. 2. New York: McGraw-hill, 1968.
16. Roddier, François. "V The Effects of Atmospheric Turbulence in Optical Astronomy." Progress in optics 19 (1981): 281-376.
17. Hardy, John W., J. E. Lefebvre, and C. L. Koliopoulos. "Real-time atmospheric compensation." JOSA 67, no. 3 (1977): 360-369.
18. Bondy, John Adrian, and Uppaluri Siva Ramachandra Murty. Graph Theory with Applications. Vol. 6. London: Macmillan, 1976.
19. Lu, Hsueh-I., and Ramamurthy Ravi. "Approximating maximum leaf spanning trees in almost linear time." Journal of algorithms 29, no. 1 (1998): 132-141.
20. Halldórsson, Magnús M., and Jaikumar Radhakrishnan. "Greed is good: Approximating independent sets in sparse and bounded-degree graphs." Algorithmica 18, no. 1 (1997): 145-163.
21. Roddier, François. "V The Effects of Atmospheric Turbulence in Optical Astronomy." Progress in optics 19 (1981): 281-376.
22. Wang, S. C. H., and M. A. Plonus. "Optical beam propagation for a partially coherent source in the turbulent atmosphere." JOSA 69, no. 9 (1979): 1297-1304.
23. Fried, David L. "Branch point problem in adaptive optics." JOSA A 15, no. 10 (1998): 2759-2768.
24. Fried, David L., and Jeffrey L. Vaughn. "Branch cuts in the phase function." Applied Optics 31, no. 15 (1992): 2865-2882.
25. Vorontsov, Mikhail. "Adaptive photonics phase-locked elements (APPLE): system architecture and wavefront control concept." In Optics & Photonics 2005, pp. 589501-589501. International Society for Optics and Photonics, 2005.
26. Noll, Robert J. "Zernike polynomials and atmospheric turbulence." JOsA 66, no. 3 (1976): 207-211.
27. Zhou, Pu, Zejin Liu, Xiaolin Wang, Yanxing Ma, Haotong Ma, Xiaojun Xu, and Shaofeng Guo. "Coherent beam combining of fiber amplifiers using stochastic parallel gradient descent algorithm and its application." Selected Topics in Quantum Electronics, IEEE Journal of 15, no. 2 (2009): 248-256.
28. Sandalidis, Harilaos G., Theodoros A. Tsiftsis, and George K. Karagiannidis. "Optical wireless communications with heterodyne detection over turbulence channels with pointing errors." Journal of lightwave technology 27, no. 20 (2009): 4440-4445.